\documentclass[12pt]{book}
\usepackage[dvips]{graphicx,color}
\usepackage{makeidx,tsukuba}
\makeauthorindex
\makeindex

\begin{document}
\BookTitle{\itshape The 28th International Cosmic Ray Conference}
\CopyRight{\copyright 2003 by Universal Academy Press, Inc.}
\pagenumbering{arabic}
\chapter{High Energy Tau Neutrinos: Production, Propagation and Prospects of
observations}

\author{
H. Athar,$^{1,3}$ Jie-Jun Tseng,$^{2}$ and Guey-Lin Lin$^{3}$ \\
{\it (1) Physics Division, National Center for Theoretical
         Sciences, 101 Section 2, Kuang Fu Road,
         Hsinchu 300, Taiwan, E-mail: athar@phys.cts.nthu.edu.tw\\
(2) Institute of Physics, Academia Sinica, Nankang, Taipei 115, 
    Taiwan, E-mail: gen@phys.sinica.edu.tw \\
(3) Institute of Physics, National Chiao Tung University, 1001 Ta Hsueh 
    Road, Hsinchu 300, Taiwan, E-mail: glin@cc.nctu.edu.tw} \\}

\section*{Abstract}
High energy tau neutrinos with energy greater than several 
thousands of GeV may be produced in some astrophysical sites. 
 A summary of the intrinsic high 
energy tau neutrino flux estimates from some representative 
 astrophysical sites is 
 presented including  the effects of  neutrino 
flavor oscillations. The presently envisaged prospects of 
observations of the oscillated high energy tau neutrino flux are  
 mentioned. In particular, a recently suggested possibility of 
future observations of Earth-skimming high energy tau neutrinos 
is briefly discussed.
\section{Introduction}
High energy neutrino astronomy ($E \geq 10^{3}$ GeV) holds great promise to unveil 
the microscopic details of the hitherto unexplored phenomenons occurring in the cosmos
around us [1,2]. In particular, a search for high energy tau neutrinos will not only bring
unparalleled new information about the cosmos around us but
 also will indicate the existence of physics beyond the 
 standard model of particle physics, namely the 
 corroboration of 
the neutrino flavor mixing [3]. 
 The essentially maximal 
neutrino flavor mixing implies the existence of high energy 
tau neutrino flux {\tt comparable} to high energy muon neutrino flux
[2]. 
\section{High energy tau neutrinos}
\subsection{Production}
The high energy tau neutrinos are produced in direct and 
 indirect decays of $D_{S}$ meson in $p(\gamma ,p)\to D_{S}+X$
 considered to be occurring in cosmos around us.
Here, the first $p$ represents the accelerated cosmic rays. 
However, the intrinsic high energy tau neutrino flux is 
 rather suppressed during its production relative to intrinsic high 
 energy non tau neutrino flux in above interactions,
 such as the intrinsic muon neutrino flux produced in the 
 atmosphere of Earth in $pA$ interactions. 
\subsection{Propagation}
Neutrino flavor mixing seems to play a dominant role 
 during the propagation of a system 
 of mixed high energy 
neutrinos over cosmologically large distances. 
\begin{figure}[t]
  \begin{center}
    \includegraphics[height=15.5pc]{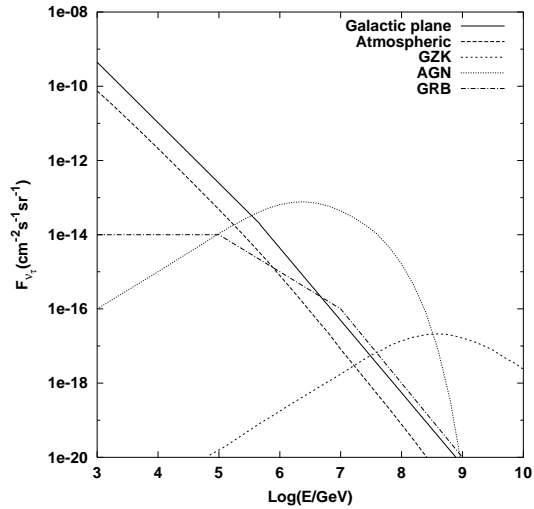}
  \end{center}
  \vspace{-0.5pc}
  \caption{Galactic plane, horizontal atmospheric 
 and GZK high energy tau neutrino flux under the
 assumption of neutrino flavor mixing. For comparison,
 the oscillated high energy tau neutrino flux 
 in a representative AGN model and in a fireball model 
 of GRB  is also shown.}
\end{figure}
Fig. 1 summarizes the intrinsic high energy tau neutrinos flux,
 $F_{\nu_{\tau}}\equiv {\rm d}N_{\nu_{\tau}}/{\rm d}({\rm log_{10}}E)$ 
 estimates from some representative astrophysical 
 sites including the effect of (two) 
 neutrino flavor oscillations [4].

A main background in observing high energy muon 
neutrinos from the cosmos is the atmospheric 
muon neutrino background. From Fig. 1, we see that 
 this is {\tt not} the case 
for high energy tau neutrinos as the
oscillated high energy tau neutrino flux from the
galactic plane is already above the relevant atmospheric 
 tau neutrino background even for $E$ as low as $10^{3}$ GeV. 
 Thus, above around $10^{3}$ GeV, 
 the high energy {\tt tau neutrino} search can lead to 
 identification of 
 extra-atmospheric neutrino flux, in particular 
 from the direction of galactic plane. This is in 
contrast to the situation for non tau neutrinos, 
where the atmospheric background tend to dominate
up to $10^{5}$ GeV for the same astrophysical site [5].  
 Thus, at least approximately 
a two order of magnitude {\tt lower} energy {\tt tau} neutrinos 
can probe same astrophysical site provided relevant tau  
 neutrinos can be isolated from relevant non tau 
 neutrinos in near future. 

\section{Prospects of possible observations}
\subsection{Downward going}
For $E\geq 5\cdot 10^{5}$ GeV, the 
downward going high energy tau neutrinos 
can be separated from other two neutrino 
 flavors via event topology characterization
 in water or ice based high energy neutrino 
 telescopes. 
It thus might become possible to identify the 
 downward going high energy tau neutrinos 
through double shower technique. The first shower is 
 from deep inelastic charged current high energy
 tau neutrino-nucleon interaction, whereas the
 second shower is from hadronic decay of the
 associated tau lepton, both showers 
 are considered to occur inside the telescope.  Some 
 quantitative  details of this idea for under 
 construction km$^{3}$ 
 ice or water based high energy neutrino telescopes are
given in [6]. 
\subsection{Upward going}
High energy tau neutrinos that cross the 
entire Earth or most part of it before
reaching the detector
are called upward going high energy neutrinos.
\begin{figure}[t]
  \begin{center}
    \includegraphics[height=15.5pc]{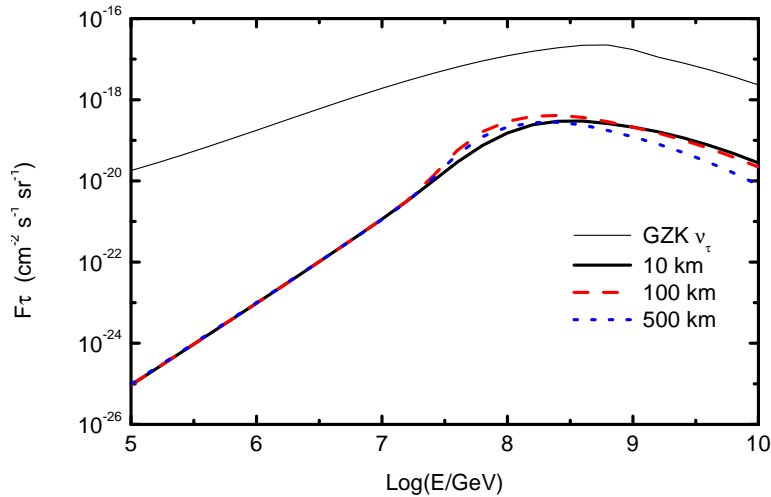}
  \end{center}
  \vspace{-0.5pc}
  \caption{The tau lepton energy spectrum induced by the rock-skimming GZK neutrinos for three
 different slant depth/matter density ratio values (see text for some more details). 
 The incident tau neutrino
 flux is shown by thin continuous line.}
\end{figure}
In the case of tau neutrinos, for $E\geq 5\cdot 10^{4}$ GeV,
 the upward going tau neutrino flux is  exponentially
suppressed [7,8]. 
\subsection{Quasi horizontal or Earth-skimming}
It might become possible in near future that 
the dedicated high energy neutrino telescopes 
 can be configured such that
the Earth-skimming
high energy (tau) neutrinos induced charged leptons as well as their
 associated radiations can be measured [9].

Figure 2 gives the tau lepton flux 
 induced by the Earth-skimming high energy 
 GZK tau neutrinos as a representative example. 
 This tau lepton energy 
spectrum is obtained by solving a 
coupled set of partial differential 
equations describing the simultaneous 
 propagation of tau neutrinos and their induced tau leptons
 inside the Earth.  
This set of differential equations 
explicitly take into account the
{\tt inelasticity} of the neutrino-nucleon
scattering and the tau lepton {\tt energy
 loss} in contrast to previous relevant 
studies in this context [10]. 
 The high energy GZK tau neutrino flux 
 is given in Fig. 1. 
 The presence of rather small tau lepton pile up around 
 $10^{8}$ GeV is a consequence
of specific energy dependence of  incident GZK high energy 
tau neutrino flux (see Fig. 1)[11].

\section{Conclusions}
\begin{itemize}
\item
The high energy tau neutrino flux can 
probe the cosmos around us such as
 the center of our galaxy even with energy 
starting from $10^{3}$ GeV {\tt above} the
 atmospheric high energy tau neutrino
 flux as a result of neutrino flavor oscillations.
This is in {\tt contrast} to the case for 
high energy muon neutrino flux.

\item
The energy spectrum of the
 tau leptons induced by the Earth-skimming high energy tau
neutrinos  is calculated by explicitly taking 
 into account the {\tt inelasticity} of the neutrino-nucleon interactions
 and the tau lepton {\tt energy loss}.
\end{itemize}
 
 H. A.  thanks Physics Division of 
National Center for Theoretical Sciences
for support. J. -J. Tseng and G. -L. Lin are
supported by the National Science Council of Taiwan under the
grant numbers NSC91-2112-M-009-019 and NSC91-2112-M-009-024.
\vspace{\baselineskip}
\re
1.\ Athar H. 2002, arXiv:hep-ph/0209130
\re
2.\ Athar H. 2002,
arXiv:hep-ph/0212387
\re
3.\ Athar H., Cheung K., Lin G. -L., Tseng J. -J.
 2003,  Astropart.\ Phys.\ 18, 581 
\re
4.\ Athar H. 2002,
arXiv:hep-ph/0210244.
\re
5.\ Protheroe R. J. 1999,
Nucl.\ Phys.\ Proc.\ Suppl.\  77, 465 
\re
6.\ Athar H., Parente G., Zas E. 2000,
Phys.\ Rev.\ D  62, 093010 
\re
7.\ Halzen F., Saltzberg D.\ 1998, 
Phys.\ Rev.\ Lett.\  81, 4305 
\re
8.\ Beacom J. F., Crotty P., Kolb E. W. 2002,
Phys.\ Rev.\ D 66, 021302 
\re
9.\ Fargion D. 2002,
Astrophys.\ J.\  570, 909
\re
10.\ Feng J. L., Fisher P., Wilczek F., Yu T. M. 2002,
Phys.\ Rev.\ Lett.\ 88, 161102 
\re
11. Tseng. J. -J. et al. 2003, arXiv:astro-ph/0305507 (to appear in Phys. Rev. D) 

\endofpaper
\end{document}